\documentclass[longauth,letter]{aa} % for the letters 
\usepackage{graphicx}
\usepackage{ae,lmodern} % ou seulement l'un, ou l'autre, ou times etc.
\usepackage{txfonts}
\usepackage{ulem}
\usepackage[]{xcolor}
\usepackage[colorlinks,allcolors=blue]{hyperref}
\begin{document} 

%
%   \title{Dynamical Interaction in the young UX Tauri multiple %system. Evidence for an on-going Fly-By.}
      \title{Ongoing flyby in the young multiple system UX Tauri}
   \titlerunning{Ongoing flyby in UX Tau}
   \author{F. M\'enard\inst{1}
\and N. Cuello\inst{1,2,3}
\and C. Ginski\inst{4,5}
\and G. van der Plas\inst{1}
\and M. Villenave\inst{1,6}
\and J.-F. Gonzalez\inst{3}
\and C. Pinte\inst{7,1}
\and M. Benisty\inst{1}
\and A. Boccaletti\inst{8} 
\and D.J. Price\inst{7}
\and Y. Boehler\inst{1}
\and S. Chripko\inst{9}
\and J. de Boer\inst{5}
\and C. Dominik\inst{4}
\and A. Garufi\inst{10}
\and R. Gratton\inst{11}
\and J. Hagelberg\inst{12}
\and Th. Henning\inst{13}
\and M. Langlois\inst{3}
\and A.L. Maire\inst{14,13}
\and P. Pinilla\inst{13}
\and G.J. Ruane\inst{15}
\and H.M. Schmid\inst{16}
\and R.G. van Holstein\inst{5,6}
\and A. Vigan\inst{17}
\and A. Zurlo\inst{17,18,19}
\and N. Hubin\inst{20}
\and A. Pavlov\inst{13}
\and S. Rochat\inst{1}
\and J.-F. Sauvage\inst{21,17}
\and E. Stadler\inst{1}
}
\authorrunning{M\'enard et al.}
\institute{Univ. Grenoble Alpes, CNRS, IPAG, F-38000 Grenoble, France\\
              \email{francois.menard@univ-grenoble-alpes.fr}
\and
Instituto de Astrof\'isica, Pontificia Universidad Cat\'olica de Chile, Santiago, Chile
\and
Univ Lyon, Univ Claude Bernard Lyon1, Ens de Lyon, CNRS, Centre de Recherche Astrophysique de Lyon UMR5574, F-69230, Saint-Genis-Laval, France
\and
Sterrenkundig Instituut Anton Pannekoek, Science Park 904, 1098 XH Amsterdam, The Netherlands
\and
Leiden Observatory, Leiden University, PO Box 9513, 2300 RA Leiden, The Netherlands
\and 
European Southern Observatory, Alonso de C\'ordova 3107, Casilla 19001, Vitacura, Santiago, Chile
\and
School of Physics and Astronomy, Monash University, Clayton Vic 3800, Australia
\and
LESIA, Observatoire de Paris, Universit\'e PSL, CNRS, Sorbonne Universit\'e, Univ. Paris Diderot, Sorbonne Paris Cit\'e, 5 place Jules Janssen, 92195 Meudon, France
\and
CECI, Université de Toulouse, CNRS, CERFACS, Toulouse, France
\and
INAF, Osservatorio Astrofisico di Arcetri, Largo Enrico Fermi 5, I-50125 Firenze, Italy
\and 
INAF, Osservatorio Astronomico di Padova, Vicolo dell'Osservatorio 5, 35122, Padova, Italy
\and
Geneva Observatory, University of Geneva, Chemin des Mailettes 51, 1290 Versoix, Switzerland
\and
Max Planck Institute for Astronomy, K\"onigstuhl 17, D-69117 Heidelberg, Germany
\and
STAR Institute, Universit\'e de Li\`ege, All\'ee du Six Ao\^ut 19c, B-4000 Li\`ege, Belgium
\and
Jet Propulsion Laboratory, California Institute of Technology, 4800 Oak Grove Dr., Pasadena, CA 91109, USA
\and
Institute for Particle Physics and Astrophysics, ETH Zurich, Wolfgang-Pauli-Strasse 27, 8093 Zurich, Switzerland
\and
Aix Marseille Universit\'e, CNRS, CNES,  LAM, Marseille, France
\and
N\'ucleo de Astronom\'ia, Facultad de Ingenier\'ia y Ciencias, Universidad Diego Portales, Av. Ejercito 441, Santiago, Chile
\and
Escuela de Ingenier\'ia Industrial, Facultad de Ingenier\'ia y Ciencias, Universidad Diego Portales, Av. Ejercito 441, Santiago, Chile
\and
European Southern Observatory (ESO), Karl-Schwarzschild-Str. 2, 85748 Garching, Germany
\and
DOTA, ONERA, Université Paris Saclay, F-91123, Palaiseau France
}
   \date{Received ; accepted }

\abstract{
We present observations of the young multiple system UX~Tauri to look for circumstellar disks and for signs of dynamical interactions. We obtained SPHERE/IRDIS deep differential polarization images in the J and H bands. We also used ALMA archival CO data. Large extended spirals  are well detected in scattered light coming out of the disk of UX~Tau~A. The southern spiral forms a bridge between UX~Tau~A and C. These spirals, including the bridge connecting the two stars, all have a CO (3-2) counterpart seen by ALMA. The disk of UX Tau C is detected in scattered light. It is much smaller than the disk of UX~Tau~A and has a major axis along a different position angle, suggesting a misalignment. We performed {\sc Phantom} SPH hydrodynamical models to interpret the data. The scattered light spirals, CO emission spirals and velocity patterns of the rotating disks, and the compactness of the disk of UX~Tau~C all point to a scenario in which UX~Tau~A has been perturbed very recently ($\sim$~1000 years) by the close passage of UX~Tau~C.
}
   \maketitle
% -----------------------
\section{Introduction}
\label{sec:intro}
Star formation occurs in molecular clouds where the stellar density is higher than in the field and the probability for encounters and dynamical interactions is enhanced. The presence of other stars in the vicinity of a forming young stellar and planetary system can dramatically affect the disk morphology and evolution \citep{Pfalzner2003, Vincke2015, Bate2018}.

The probability of a system to undergo a flyby decreases rapidly with time in a stellar association in unison with the stellar density, which decreases with cluster expansion. During the first million years of a stellar cluster, \citet{Pfalzner2013} and \citet{Winter2018} estimated that the probability of a stellar encounter can be on the order of 30\% for solar-type stars, a flyby being defined in this case as a single passage within 100-1000~au.
These calculations assumed a background stellar density that is larger than in Taurus. In Taurus the stellar density is low (1-10 stars / pc$^{3}$) and at first sight the flyby rate would be equivalently low. However, the stellar distribution in Taurus is patchy; several denser groups have been identified in Taurus \citep[e.g.,][]{Joncour2018} and there is a higher probability that encounters might happen in these groups.

\citet{Clarke1993} considered coplanar parabolic encounters between equal-mass stars with periastron separations on the order of the initial disk size and found that prograde encounters were the most destructive. These prograde encounters tidally truncate the disk and unbind material that is either captured by the perturber or escapes \citep{Breslau2017}. Retrograde encounters, on the contrary, were found to be much less perturbative. More recently, \citet{Cuello2019} showed through hydrodynamical simulations -- including dust and gas -- that the solids within the disk react differently than the gas to the perturbation because of gas drag and radial drift. This dynamical effect renders the gas spirals rich in micron-sized grains (well coupled to the gas) and hence detectable in scattered light. In addition, the disk seen in thermal emission is expected to be more compact since larger grains are more prone to radial drift \citep{Weidenschilling1977}. The detailed observational signatures of these flyby models -- obtained through radiative transfer post-processing -- can be found in \citet{Cuello2020}.

Only a few cases of flybys have been observed so far: RW Aur \citep{Dai2015}, HV~Tau and DO~Tau \citep{Winter2018b}, and AS~205 \citep{Kurtovic2018}. \citet{Zapata2020} recently suggested that UX~Tauri might also be the site of dynamical interactions. In this Letter we present SPHERE images of the UX Tauri multiple system. Supported by tailored hydrodynamical simulations, the new data strongly favors a recent passage of UX Tauri C near UX Tauri A, whose disk is still being perturbed.

The UX~Tauri system includes four T~Tauri stars. It is located in the Taurus molecular cloud at d = $147\pm2$~pc \citep{GAIA-DR2-2018}. This system consists of a primary star (UX~Tau~A) and two companions: UX~Tau~B at $\sim5\farcs8$ to the west and UX~Tau~C at $\sim2\farcs7$ to the south of
the primary. UX~Tau~B is itself a tight ($\sim$ 0\farcs1) binary
%system whose orbit has been followed by several authors
\citep[e.g.,][]{Duchene1999,Correia2006,Schaefer2014}.  \citet{Kraus2009} estimated the stellar masses at 1.3$\pm 0.4$~M$_{\odot}$ for A and 0.16$\pm0.04$~M$_{\odot}$ for C.
\citet{Zapata2020} report dynamical masses from CO data: 1.4$\pm$0.6 M$_\odot$ for A, and (0.067$\pm 0.029 / \sin{i})$ M$_{\odot}$ for C.
%The spectral type of UX~Tau~A is K2 %\citep{Kraus2009}. 
So far only the disk of UX~Tau~A has been detected. 
\citet{Andrews2011}, using the Submillimeter Array (SMA), resolved an inner cavity around the central star extending to a radius of 25~au: the dust disk is a ring at 880~$\mu$m. \citet{Tanii2012} resolved the disk at H band with SUBARU/HiCIAO in polarized intensity and detected scattered light emission all the way to the coronagraphic mask, that is, down to r = 0\farcs15 or r = 22~au. Unfortunately, the coronagraphic mask has a size similar to the millimeter-emission cavity and it is not known whether small dust (detectable in scattered light) is present in the millimeter cavity or not. The shape of the spectral energy distribution however led \citet{Espaillat2007} to identify UX~Tau~A as a so-called pre-transition disk, that is, a disk that shows a slight near-infrared (NIR) excess and significant excesses in the mid- and far-infrared. Interestingly, a SPITZER IRS spectrum revealed the absence of any silicate feature emission at 10~$\mu$m \citep{Espaillat2010}, also implying a lack of small hot silicate dust grains in the disk.

In \S\ref{sec:observations} we present new SPHERE high-contrast coronagraphic images and archival ALMA data. In \S\ref{sec:hydro} we present hydrodynamical models of the gravitational interaction in the UX Tauri A-C system. In \S\ref{sec:discuss} we discuss the data and model results and their implications for the disk structure and evolution. In \S\ref{sec:conclusion} we summarize our findings.

%-------------------------
\section{Observations and results}
\label{sec:observations}

\subsection{Near-infrared high-contrast imaging with SPHERE}
\begin{figure}[htb]
    \centering
\includegraphics[width=0.92\columnwidth]{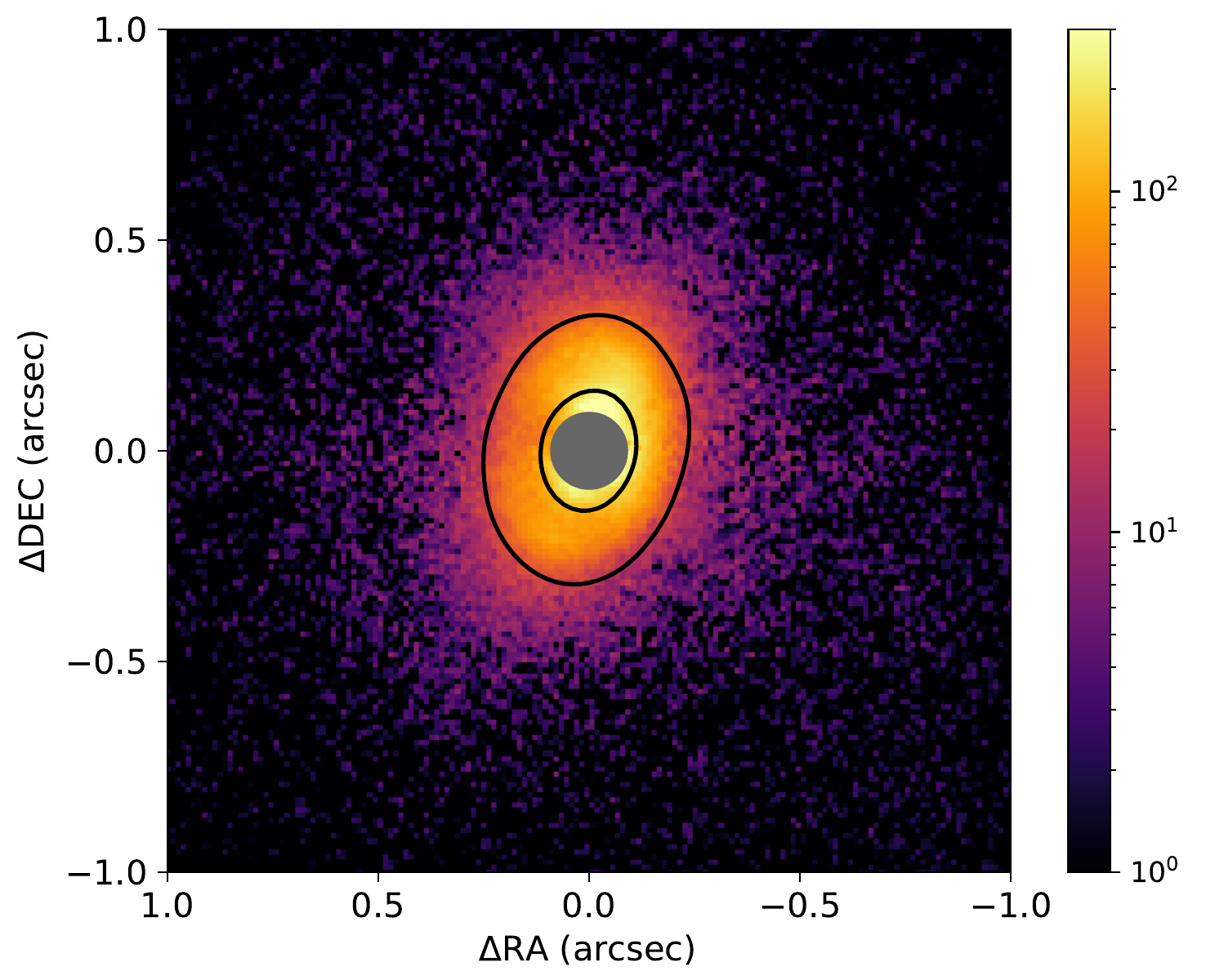}
    \caption{Q$\phi$,  J-band (linearly) polarized intensity image of the disk of UX~Tau~A, shown in logarithmic scale. The intensity scale is in arbitrary detector units, levels below 1 (i.e., $\sim$2.5 times the rms noise level at $\geq$1") are set to black.   The gray circle traces the 185 mas coronagraphic mask. The continuum ring detected by ALMA (see fig.~\ref{fig:ALMAco}, left insert) is shown in black contours traced at 60\% of the ALMA continuum peak level.}
    \label{fig:UX-ovrl}%
\end{figure}
\begin{figure}[t]
    \centering
   \includegraphics[width=0.92\columnwidth]{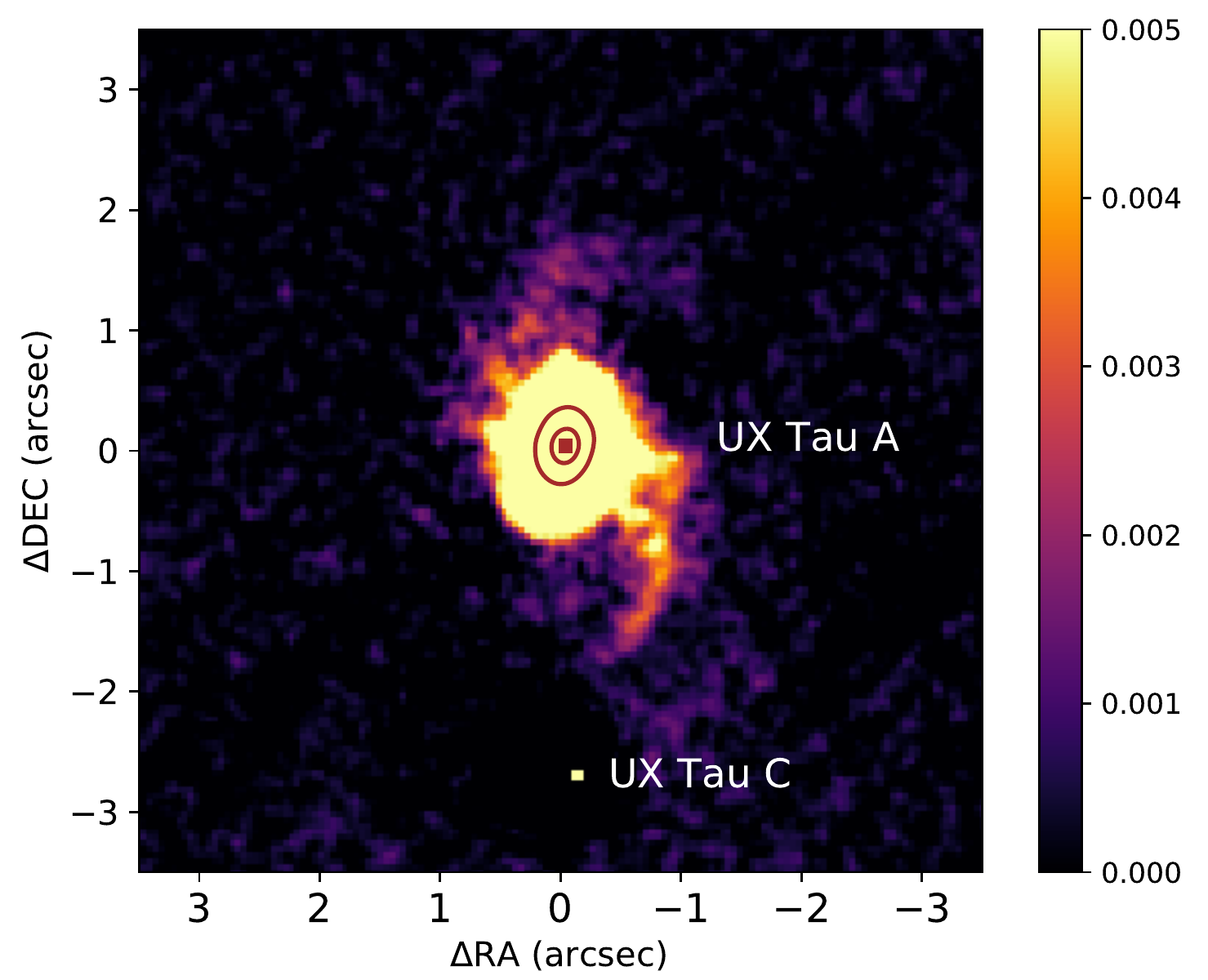}
    \caption{J- and H-band combined image. The native image {\bf(12.25 mas/px)} was binned by a factor of 4 and smoothed with a Gaussian filter of 2 binned pixels. The intensity scale is relative to the peak of the image. The color scale saturates (in yellow) at 0.5\% of the peak. The mean signal in the southern spiral is detected at 3.9$\sigma$ above background in the binned images, before smoothing. For comparison, the dust ring seen by ALMA is overplotted in red, to scale. A different rendering of the spirals in the J band, using a novel de-noising procedure (Price et al. 2020, in prep.), is presented fig.~\ref{fig:denoised}}
    \label{fig:deepAO}%
\end{figure}
We obtained three different NIR  differential polarization imaging data sets (DPI; see \citet{vanHolstein2020,deBoer2020}) with the IRDIS camera \citep{Dohlen2008} of SPHERE \citep{Beuzit2019}. Two data sets were obtained in the broadband H filter (as part of the GO program 0100.C-0452; PI Benisty) and one was taken in the J band (as part of the SPHERE DISK GTO program 096.C-0248). The H-band data were taken on 2017 October 6 and 12, the J-band data on 2015 December 18. The final data reduction was performed with the IRDAP pipeline\footnote{\url{https://irdap.readthedocs.io}} presented in \citet{vanHolstein2020}. The data reduction process is described in detail in the same paper. 

The disks of UX~Tau~A and UX~Tau~C are both resolved in the J and H bands. Figure~\ref{fig:UX-ovrl} shows the J-band polarized intensity image of UX~Tau~A in color, with the ALMA continuum ring overlayed in white contours. The SPHERE image shows that the disk is brighter on the western side, suggesting this is the front side, facing us. This is a natural consequence of the forward-scattering phase function of the grains at the disk surface. This implies, coupled with the velocity gradient presented in \S\ref{sec:alma}, that the disk is rotating counter-clockwise as seen by the observer. The compact disk of UX~Tau~C is shown in Fig.~\ref{fig:UXC-Jband}.

Figure~\ref{fig:deepAO} shows a deep rendering of the close environment of UX~Tau~A. The disk is the bright saturated yellow patch at the center. Two spirals are clearly visible emerging from the disk: one toward the north and one toward the south. The locations of UX~Tau~A and C are indicated. UX~Tau~C has a projected separation of 2.70\arcsec$\pm$0.02\arcsec at PA = 181.5\degr$\pm$1.0\degr (east of north; this is the convention throughout the paper). 
The southern spiral extends toward UX~Tau~C, creating a bridge between UX~Tau~A and C, which is immediately suggestive of a dynamical interaction. We note, critically, that the two spirals are detected in each individual data set, although at lower signal-to-noise ratio (S/N).

\subsection{ALMA archival data}
ALMA CO observations of UX Tauri were presented in \citet{Francis2020} and \citet{Zapata2020}. We used the same data set. The data reduction procedure and the main results are presented in Appendix~\ref{sec:alma}. We highlight the main results in Fig.~\ref{fig:COmom8}. Disks are detected around both stars and a comparison with Fig.~\ref{fig:deepAO} immediately reveals the correspondence between the scattered light and CO spirals. This is discussed in \S~\ref{sec:flyby}
\section{Hydrodynamical modeling of UX Tau}
\label{sec:hydro}

We now test the hypothesis that the spirals observed UX~Tau~A are due to the dynamical response of the disk to a gravitational perturbation by UX~Tau~C. In the following, we place ourselves in the reference frame of UX~Tau~A and call UX~Tau~C the perturber.

We model the hydrodynamical evolution of the disk around UX Tau A by means of the smoothed particle hydrodynamics (SPH) {\sc Phantom} code \citep{Price2018b}. The disk is modeled using 500\,000 gas SPH particles and assuming a total disk mass of $0.05~M_\odot$ \citep{Akeson2019}. At the beginning of the calculation, the disk surface density follows a power-law profile $\Sigma \propto R^{-1}$. We further assume that the disk is vertically isothermal. 
We adopt a mean Shakura–Sunyaev disk viscosity $\alpha_{\mathrm SS} \approx 0.005$ by setting a fixed artificial viscosity parameter $\alpha_{\mathrm AV}$ = 0.25 and using the 'disk viscosity' flag of {\sc Phantom} \citep{Lodato2010}. The disk inner and outer radii are initially set to $R_\mathrm{in}=10$~au and $R_\mathrm{out}=60$~au. The value of $R_\mathrm{in}$ -- although in agreement with the SPHERE and ALMA observations -- is not critical in this study since we are interested in the spirals external to the disk. The value of $R_\mathrm{out}$ is more difficult to choose a priori because we expect the disk to be truncated as a result of the gravitational interaction with the perturber. Therefore $R_\mathrm{out}$ should not be compared directly with the observations. In order to ease the comparison with the observations, the disk is set with an inclination $i_\mathrm {d}=40\degr$ and PA$=167\degr$ before the perturbation, following the analysis of \citet{Francis2020}.

There are few observational constraints regarding the orbit of the perturber. This yields a highly degenerate geometry for the multiple system. The projected separation in the plane of the sky between UX Tau C and UX Tau A is known: d$_\mathrm{C}\approx400$~au at position angle $\theta_\mathrm{C}=181\degr$, east of north. There is no apparent motion between A and C in the period 2009-2018 ($<$0\farcs05, 7au). Lastly, the mass ratio $q=M_\mathrm{C}/M_\mathrm{A}$ ranges between 0.08 and 0.22 \citep{Kraus2009,Zapata2020}. The systemic velocities measured by \citet{Zapata2020} indicate, marginally, that UX Tau C is slowly approaching to us with respect to UX Tau A \citep{Zapata2020}. 
In this work, we do not intend to explore the whole orbital parameter space available, but rather to compare two possible scenarios: a bound companion where the orbital eccentricity $0 \leq e<1$ and a parabolic flyby where $e=1$. These runs are labeled B- and F-, respectively. We assume that the orbit of the perturber lies in the plane of the sky. Hence there is an angle of $40\degr$ between the orbit of the perturber and the disk midplane. The masses of the primary star and the perturber are set to 1.0 M$_\odot$ and 0.2 M$_\odot$, for a mass ratio $q=0.2$, within the range given above.
\begin{figure*}
    \centering
   \includegraphics[width=0.9\textwidth]{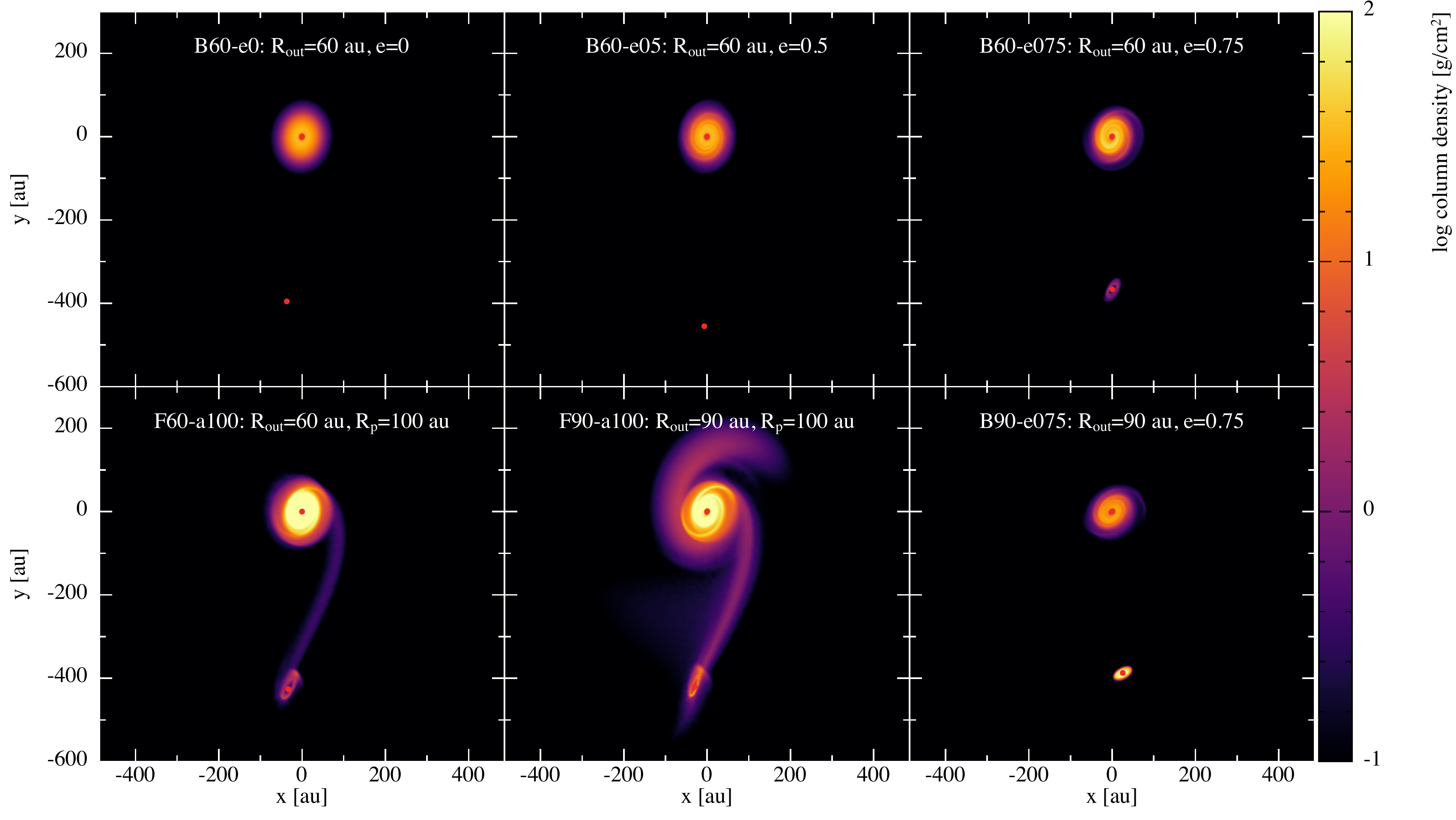}
    \caption{Gas column density of the disk around UX~Tau~A being gravitationally perturbed by UX~Tau~C. The orbit of the perturber is in the plane of the sky and the disk is inclined by $40\degr$ with PA$=167\degr$, as in the observations. The letters F- and B- label flyby and bound orbit models, respectively. The morphology of the spirals and the current projected distance between the stars suggest that UX~Tau~C is on an unbound orbit. 
%An (eccentric) companion on a bound orbit is unable to reproduce the observations. 
We note that increasing the outer disk radius for a flyby (as in F90-a100) renders the spirals denser and brighter in scattered light, as they capture more disk material. In each panel, the position (separation, PA) and relative motion (proper motion, relative V$_\mathrm{rad}$) of UX Tau C are compatible with observations. Videos of F90-a100 and B90-e075 are available online.}
    \label{fig:sims}
\end{figure*}

For B-runs, we fix the semimajor axis to $400$~au, that is, equal to the current projected distance, and vary the orbital eccentricity: $e=0$ (B60-e0), $e=0.5$ (B60-e05), and $e=0.75$ (B60-e075). In addition, we perform a calculation with $R_\mathrm{out}=60$~au and $e=0.75$ (B90-e075) as the most extreme bound orbit. For $e=0.5$ and $e=0.75$, the distance between the stars at pericenter, $R_\mathrm{p}$, is equal to 200 and 100~au, respectively. The morphology of the spirals and the location of the perturber in the images suggest that the latter is currently past pericenter \citep{Cuello2019}. We find that an argument of pericenter of $\omega=-130\degr$ simultaneously produces a pair of spirals (shortly after pericenter passage) and the correct current projected separation of 400~au at position angle $\theta_\mathrm{C}=181\degr$. In the plane of the sky, the pericenter is located at position angle $\theta_\mathrm{P}=45\degr$. We compute the disk evolution for ten binary orbits. Lower values of $R_\mathrm{p}$ produce a disk truncation that is incompatible with the observations. 

For F-runs, we consider prograde flybys with $R_\mathrm{p}=100$~au and $\omega=-130\degr$ as well. The flybys are non-penetrating since $R_\mathrm{p}/R_\mathrm{out}>1$. However, because this ratio is close to 1, the flybys trigger a pair of prominent trailing spiral arms, in agreement with the observational data. For comparison, non-penetrating retrograde flybys with $q<1$ are unable to produce such prominent spiral arms \citep[e.g.,][]{Cuello2019}. Since the perturbation that truncates the disk occurs only once, the disk size prior to the encounter can in principle be larger than the observed value. Hence, we compute two simulations, F60-a100 and F90-a100, where $R_\mathrm{out}=60$ and $90$~au. The results, which are presented in Fig.~\ref{fig:sims} and discussed below, strongly favor a stellar flyby over a bound orbit in UX~Tau.

\section{Discussion}
\label{sec:discuss}

\subsection{UX Tau C: Bound companion or flyby}
\label{sec:flyby}

The results of the SPH calculations demonstrate that in B-runs the disk around UX~Tau~A always becomes heavily truncated after a few orbital periods. The direct observational consequence is that the spirals triggered by UX~Tau~C would be more tightly wound and less dense compared to those triggered by a flyby. Moreover, since in F-runs the gravitational interaction occurs only once, there is more material available at large disk radii to feed the spirals. During such a close and inclined prograde flyby, a bridge of material is created and remains between both stars for a long time after the passage at pericenter. This bridge should be detectable in deep NIR scattered light observations and deep CO line observations as shown by \citet[][see their Fig.~4]{Cuello2020}. This is because micron-sized dust grains are strongly coupled to the gas and are efficiently trapped and dragged along the gas spiral arms.

We observe that in F60-a100 and F90-a100 there is a small fraction of the initial disk that is captured by the perturber. This is an expected dynamical outcome of prograde flybys when $R_\mathrm{p}$ is close to $R_\mathrm{out}$ \citep{Cuello2019}. The captured material is expected to form a disk around the perturber, presumably on an inclined orbital plane compared to the primary disk midplane. This is also seen for B90-e075 and B60-e075 since some material is stripped from the disk around UX~Tau~A during the first orbit. However, after a few orbits the spirals become much more tenuous and the bridge disappears.

Interestingly, the velocity gradients observed in the two disks are not along the same position angle (see also \citealt{Zapata2020}) and the inclination of UX Tau C, $i \geq 60\degr$ (see Appendix~\ref{sec:suppfig}), is significantly different from UX Tau A. The two disks are likely not coplanar. In addition, the peak velocity map (Fig.~\ref{fig:ALMAco}, right panel) shows that the bridge connecting the two stars overlaps significantly with the scattered light spirals of Fig.~\ref{fig:deepAO}. This kind of kinematics and material exchange is typical of prograde flybys. The presence of a disk around the perturber prior to the flyby cannot be excluded. However, if present, that disk probably had a compact radial extension. Otherwise, given our choice of pericenter, there would be much more disk material scattered all around and the disk of UX~Tau~C would harbor large spirals. Our simulations are unable to reproduce the outer spiral located to the southwest (see right panel in Fig.~\ref{fig:ALMAco}).

The relative orientations of the disks cannot be used safely to infer the origin of the multiple system. If UX~Tau~C  acquired its disk during the flyby, then the disk misalignment is a natural outcome of the three-dimensional interaction \citep{Bate2018,Cuello2020}. If UX~Tau~A, B, and C were born from the same molecular core, misaligned disks could indicate a turbulent formation process \citep{Offner2016,Bate2018} or, more likely, trace the effects of dynamical evolution within the system. Indeed, the perturber is unlikely to come from the larger, low stellar density, Taurus star formation region,  but instead is likely to come from within the smaller UX Tau group (already known to contain three other T Tauri stars). A similar conclusion was drawn by \citet{Winter2018b} for  DO Tau and HV Tau.

\subsection{UX~Tau~A: Disk inner cavity}
At millimeter wavelengths, it is well documented that the disk of UX~Tau~A has a ringed shape \citep{Andrews2011, Pinilla2014}. \citet{Francis2020} studied in detail the continuum emission obtained from the same archival data set we used for $^{12}$CO. Our reduction of the same continuum data is presented in Fig.~\ref{fig:ALMAco}, left insert. \citet{Francis2020} modeled a decrease (a jump) in dust surface density in the cavity by a factor in excess of 200, based on the nondetection of millimeter flux inside the cavity. We show in Fig.~\ref{fig:UX-ovrl} that NIR scattered light may be marginally detected inside the cavity observed by ALMA. However, these results are not immediately in contradiction with those of \citet{Francis2020} because small dust particles have low millimeter opacity and because the coronagraphic mask used for the SPHERE observations also covers most of the cavity (see Fig.~\ref{fig:UX-ovrl}). Pending a deep non-coronagraphic image that would confirm this, the detection of scattered light inside millimeter cavities is common for transition disks whose cavities are not fully devoid of gas, as is the case here (see Fig.~\ref{fig:ALMAco} middle panel and right insert). This is also in agreement with the pre-transition disk nature identified by \citet{Espaillat2010}.

Regarding the origin of the inner cavities in transition disks, it is often suggested that they are carved by an inner very low-mass companion. One or several undetected low-mass planets could easily create a decrease in the gas surface density, thereby causing large dust particles to remain trapped at the edge of the cavity and explaining the ring shape seen by ALMA; small dust particles, however, may be able to filter into the cavity together with the accreting gas \citep[see, e.g.,][]{PerezS2015,vanderMarel2018}.  However, a more massive stellar companion within the cavity can probably be excluded on dynamical grounds since in that case we should see more asymmetries in the dust and in the gas. Similarly, a stellar companion within the cavity can potentially trigger prominent spirals arms in protoplanetary disks, as in, for example, HD~142527 \citep{Price2018}, but the radial extent and brightness of the spirals observed in UX Tau A exclude such a dynamical scenario. Therefore, based on our simulations and despite the orbital degeneracy, we conclude that the disk of UX Tau A is perturbed by a recent flyby by UX~Tau~C and that UX Tau C is likely on an unbound orbit. 

%------------------------
\section{Concluding remarks}
\label{sec:conclusion}

In this Letter we presented deep scattered light linearly polarized intensity images of the disks of UX~Tau~A and UX~Tau~C obtained with the SPHERE instrument at the ESO/VLT. Several features are revealed for the first time by SPHERE.
First, two faint and extended scattered light spirals are seen emanating from the outer disk of UX Tau A, propagating outward. The southern spiral points to and extends almost all the way to UX~Tau~C, which is very suggestive of a tidal bridge between A and C.
Second, a faint and compact disk is also resolved around UX~Tau~C. The ALMA archival $^{12}$CO (3-2) line data show unambiguously that the two spirals seen in scattered light are also detected nicely in the gas emission. The two disks of UX~Tau~A and C are also detected in CO data, with clear rotational signatures. The rotational velocity gradients of the two disks are misaligned. 

To understand the origins of the spirals, SPH hydrodynamical simulations were used to examine two hypothesis: one in which UX~Tau~C is orbiting around UX~Tau~A in a bound orbit, and one in which UX~Tau~C is a one-time perturber, having recently flown by UX~Tau~A on a parabolic orbit. A prograde flyby (e.g., panel F90-a100 in Fig.~\ref{fig:sims}) provides an excellent qualitative match to the ensemble of observations. A bound orbit solution can also provide a decent qualitative match, but only  during the first few orbits. That solution produces fainter spirals but, more problematically, requires that we are also seeing the system very early in its evolution because the spirals vanish after a few orbital periods as the disk gets truncated and the loosely bound material is no longer available, having been stripped away during previous passages. The orbital period is of order 7000 yr in our simulations. It is not expected that after 1-2 Myr, the age of UX~Tau~A, the spirals would still be formed.

In summary, the ensemble of multiwavelength observational constraints available very clearly favors the flyby scenario.
%The age of UX Tau A is of order 1-2Myr while our simulations %suggest a recent flyby.
All the reported features are also in good qualitative agreement with the dynamical and observational flyby signatures shown in \citet{Cuello2019} and \citet{Cuello2020}, in particular the prominent gaseous spirals and the compact dusty disk detected around UX Tau A. We hypothesize that the disk around UX~Tau~C was formed during the flyby shortly after the passage at pericenter. 
This would explain why the disk of UX~Tau~C is detected in CO emission (tracing gas) and in scattered light (tracing small dust), but remains undetected in the millimeter continuum (tracing larger particles). \citet[][see their \S3.2]{Cuello2020} reported that flybys in disks where radial drift has had time to sort the large from the small particles would result in the perturber capturing only gas and small dust, leaving the larger particles unaffected. This would be in good agreement with UX Tau. Given the compact size of the disk around UX~Tau~C, and its relative faintness in CO and scattered light, a deeper ALMA continuum image will be needed to confirm this conjecture. Similarly, the SPH model cannot currently explain the origin of the outermost spiral features seen in CO. No doubt, UX Tau is a fantastic laboratory to study disk dynamics during a stellar flyby that deserves further attention.

\begin{acknowledgements}
SPHERE was designed and built by a consortium made of IPAG (Grenoble, France), MPIA (Heidelberg, Germany), LAM (Marseille, France), LESIA (Paris, France), Laboratoire Lagrange (Nice, France), INAF–Osservatorio di Padova (Italy), Observatoire de Genève (Switzerland), ETH Zurich (Switzerland), NOVA (Netherlands), ONERA (France) and ASTRON (Netherlands) in collaboration with ESO.  SPHERE was funded by ESO, with additional contributions from CNRS (France), MPIA (Germany), INAF (Italy), FINES (Switzerland) and NOVA (Netherlands).  
Additional funding from EC's 6th and 7th Framework Programmes as part of OPTICON was received (grant number RII3-Ct-2004-001566 for FP6 (2004–2008); 226604 for FP7 (2009–2012); 312430 for FP7 (2013–2016)). We acknowledge the Programme National de Planétologie (PNP) and the Programme National de Physique Stellaire (PNPS) of CNRS-INSU, France, the French Labex OSUG@2020 (Investissements d’avenir – ANR10 LABX56) and LIO (Lyon Institute of Origins, ANR-10-LABX-0066 within the programme Investissements d'Avenir, ANR-11-IDEX-0007), and the Agence Nationale de la Recherche (ANR-14-CE33-0018) for support. 
We acknowledge funding from ANR of France under contract ANR-16-CE31-0013. NC acknowledges support from the European Union's Horizon 2020 research and innovation programme under the Marie Sk\l{}odowska-Curie grant agreements No 210021 and 823823. The Geryon cluster at the Centro de Astro-Ingenieria UC was extensively used for calculations. BASAL CATA PFB-06, the Anillo ACT-86, FONDEQUIP AIC-57, and QUIMAL 130008 provided funding for several improvements to the Geryon cluster.
This research has made use of the NASA Astrophysics Data System.
AZ acknowledges support from the FONDECYT Iniciaci\'on en investigaci\'on project number 11190837
\end{acknowledgements}
\bibliographystyle{aa}
\bibliography{MyBibFMe.bib}
\begin{appendix} 
\section{Disk of UX~Tau~C}
\label{sec:suppfig}
\begin{figure}[h]
   \includegraphics[width=0.95\columnwidth]{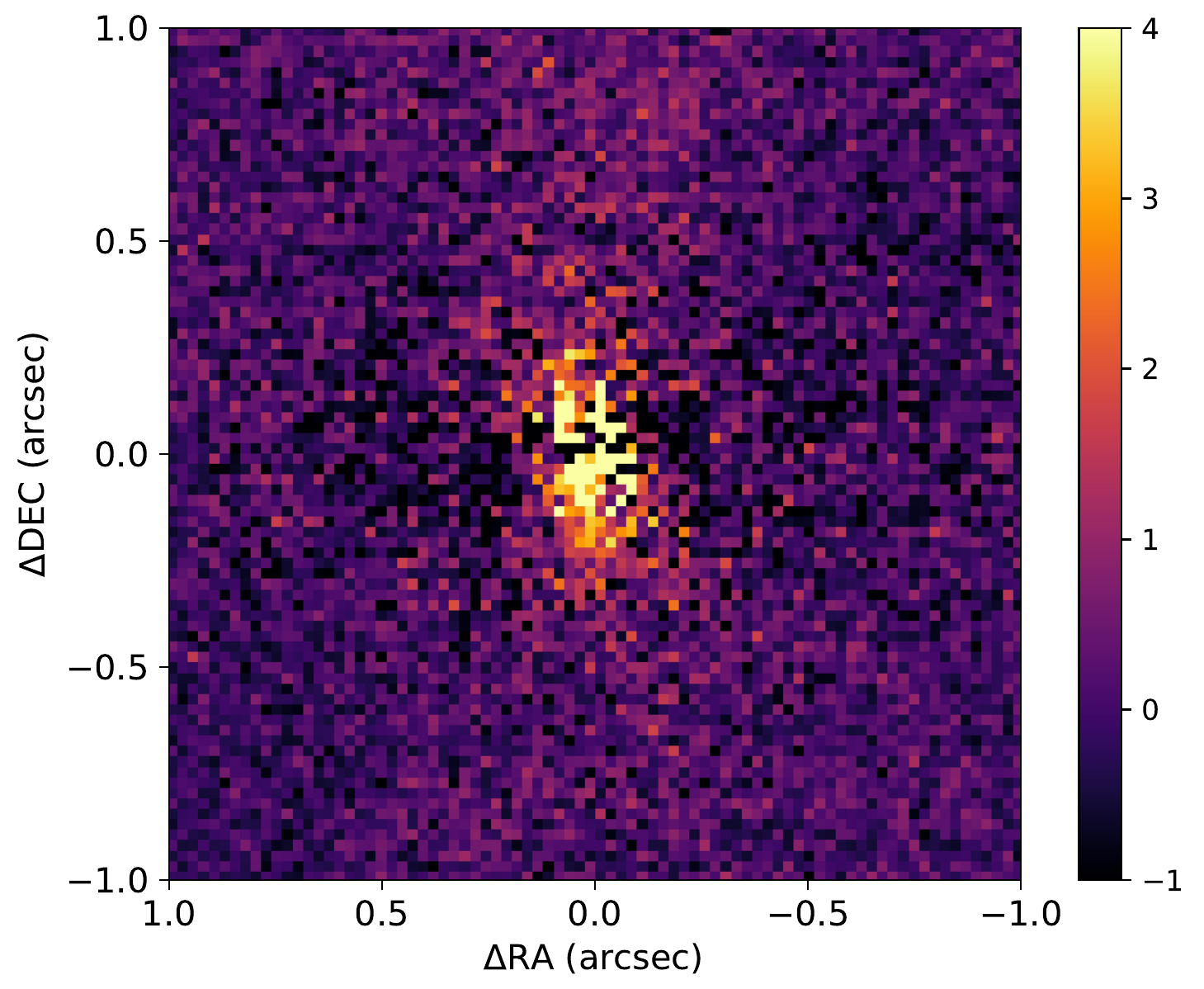}
    \caption{SPHERE J-band linearly polarized intensity image (Q$\phi$) of the disk of UX Tau C. The image is binned by a factor of 2 to improve contrast. The data are shown in linear stretch, and the scale is absolute in detector units (ADU). The disk is faint and compact. The disk position angle is PA = $5\degr \pm 2\degr$. The axis ratio suggests an inclination larger than i = 60\degr. The disk is detected over 0\farcs13, or $\sim$19~au.}
    \label{fig:UXC-Jband}
\end{figure}

\section{ALMA archival data}
\label{sec:alma}
\begin{figure}[htb]
   \includegraphics[width=\columnwidth]{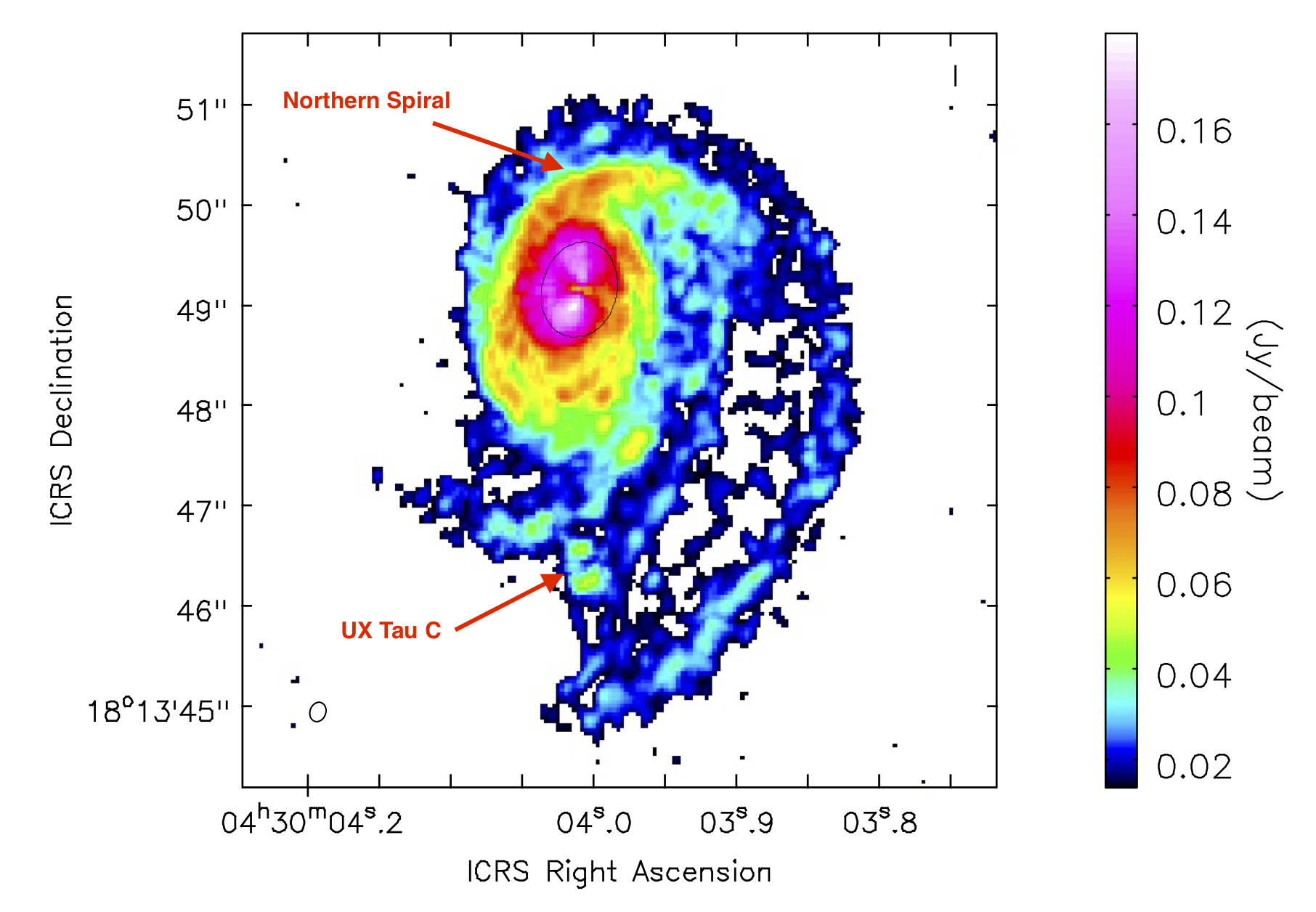}
    \caption{Peak intensity map (moment 8) made from the Briggs-weighted CO(3-2) data. The black contour is the continuum emission traced at the 1$\sigma$ level (58 $\mu$Jy). The position of UX Tau C is labeled. The northern spiral arm seen in scattered light in Fig.~\ref{fig:deepAO} is also indicated.}%
    \label{fig:COmom8}
\end{figure}
\begin{figure*}[htb]
    \centering
    \includegraphics[width=0.97\textwidth]{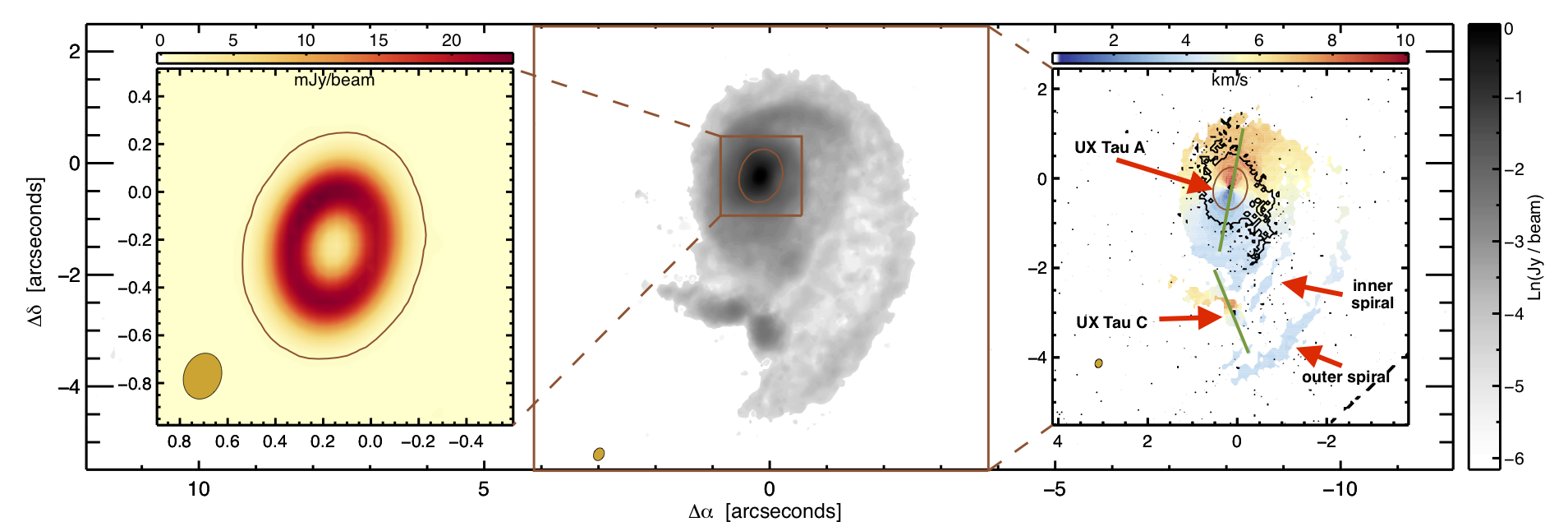}
    %contzoom_planet_2panel.pro
    \caption{CO J=3-2 integrated intensity map (moment 0, in grayscale at the center) of the region spanning UX Tau A and C, located at offset [0,0] and [0,-2.7], respectively. The left insert is a zoom of the ringed-disk seen in 0.84 mm continuum emission. The right insert is a map of the peak intensity velocity, where the rotation signature of the disk of both UX Tau A  and C are clearly seen. Labels indicating the various features are added. The two green lines indicate the direction of the velocity gradients in the disks. The deep J-band scattered light map of Fig.~\ref{fig:deepAO} is shown in black contours. In each panel the brown contour ellipse around UX~Tau~A is drawn at 10 times the continuum RMS of 58 $\mu$Jy. The beam is shown in the bottom left of each panel.}
    \label{fig:ALMAco}%
\end{figure*}

The UX~Tau system was observed multiple times with ALMA. The continuum emission of component UX~Tau~A was discussed by \citet{Pinilla2014} and \citet{Akeson2019}. Only the disk of UX~Tau~A is detected in the continuum. \citet{Francis2020} and \citet{Zapata2020} presented archival deep and high-resolution data obtained by program 2015.1.00888.S (PI Akiyama). This program included $^{12}$CO~(3-2) line emission data that we use here as well. The observations have a velocity resolution of 410 m/s for the CO line data. The observation took place on 2016 August 10. We performed our own reduction using the supplied reduction script and version 4.7.2 of the CASA software \citep{McMullin2007}. We self-calibrated the continuum visibilities using three rounds of phase-only self-calibration using successively shorter solution intervals ($\infty$, 60s, 40s), resulting in a gain of a factor $\sim$2 in RMS noise and dynamic range.

We imaged the continuum visibilities using Briggs weighting resulting in a 0\farcs20 $\times$ 0\farcs17 beam at PA $=$ -21.0\degr. For the CO emission we subtracted the continuum data using the {\sc uvcontsub} routine of CASA, and applied the self-cal solution to the resulting visibilities. We imaged the CO line using a channel width of 0.24413 MHz and a natural weighting resulting in a 0\farcs24 $\times$ 0\farcs18 beam at PA $=$ -24.7\degr. The CO line emission and the dust continuum emission maps are shown in Fig.~\ref{fig:ALMAco}. From the same continuum data, \citet{Francis2020} fit an inclination i$_{\mathrm d}=40\degr$ at a PA = 167\degr.

The middle panel of Fig.~\ref{fig:ALMAco} shows the CO (3-2) integrated intensity map (moment 0) of the region spanning UX~Tau~A and C found at offset [0,0] and [-2.7,0], respectively. The CO emission overlaps with the two stars. The left insert is a zoom of the Band 7 (0.84mm) continuum emission showing the ringed structure of the disk at that frequency. The insert on the right is an overlay of the CO (3-2) map of the velocity of the peak emission in color with the deep J-band SPHERE image of Fig.~\ref{fig:deepAO} in contours.
The bridge between components A and C suggested in Fig.~\ref{fig:deepAO} is located just inside the inner CO spiral labeled in the figure.
Two, possibly three, spirals can be identified in CO emission on the western side of UX Tau A: the one seen in scattered light, which may correspond to a distinct CO spiral; and two others labeled inner and outer spirals, which have the same general curvature but are located further away. The northern spiral seen in scattered light is also clearly detected in the CO peak intensity map shown in Fig.~\ref{fig:COmom8}. This spiral is part of the same dynamical structure creating the inner and outer CO spirals. The front face of the disk being on the western side, the faint spirals detected by SPHERE and the CO spirals observed by ALMA are trailing spirals. 

The rotational signature of the two disks is well resolved. The orientation of the velocity gradients is indicated by green lines in Fig.~\ref{fig:ALMAco}, right panel. They are not parallel to each other. SPHERE marginally resolved the disk of UX Tau C and the inclination is i$\geq$60\degr. The disks of UX Tau A and C are likely not coplanar. The tail of material attached to UX Tau C reported by \citet{Zapata2020} is also visible on the same panel, to the left of UX Tau C. All these are expected signatures of stellar flybys with pericenters approximately the size of the original primary disk radius.

\section{Novel DE-NOISING procedure}
\label{sec:denoise}

\begin{figure}[htb]
   \includegraphics[width=\columnwidth]{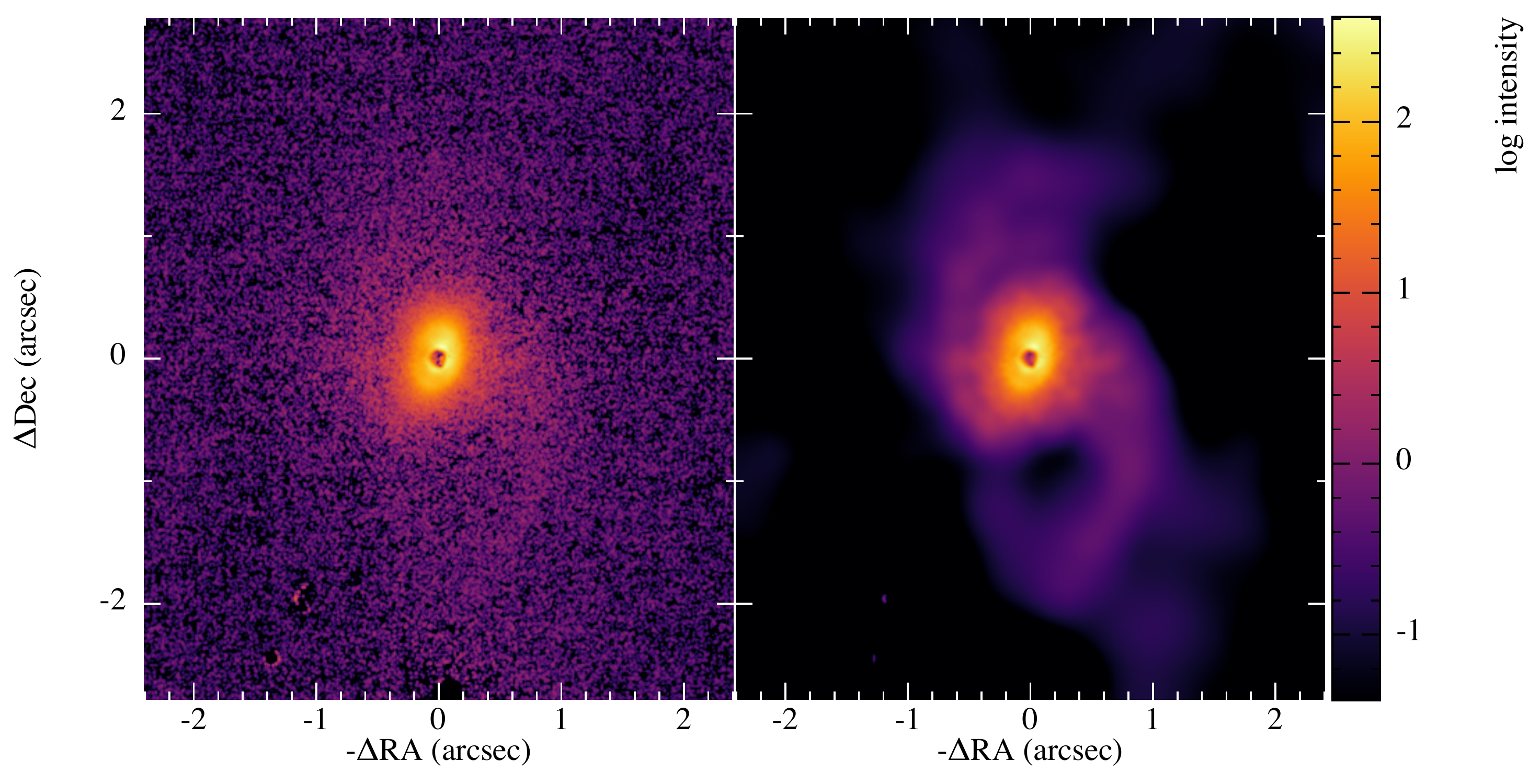}
    \caption{J-band polarized intensity images. {\bf Left panel:} Native J-band image shown in log stretch. The two spirals a barely detectable above background noise. {\bf Right panel:} Denoised J-band image. At the expense of a slight loss of angular resolution, the de-noised spirals are easily detectable.     
    }%
    \label{fig:denoised}
\end{figure}

 To smooth the noise in the SPHERE images and highlight the spirals directly from the native images without stacking or binning, we used a novel adaptive de-noising procedure based on the adaptive kernel smoothing employed in SPH and implemented in the {\sc splash} software \citep{Price2007}.  Figure~\ref{fig:denoised} shows the native J-band polarized intensity image on the left and the de-noised image in the right panel. The two spirals are visible on both images, but more contrast and more details are visible in the de-noised version. We note that only one data set was used to produce Fig.~\ref{fig:denoised}, contrarily to Fig.~\ref{fig:deepAO}.
 
 The main idea behind the de-noising procedure is to use a beam size (smoothing length) that is locally adaptive for each pixel, inversely proportional to the square root of the intensity. Specifically, we constructed a smoothed intensity image from the original image according to
\begin{equation}
<I(x,y)> = \sum_i \sum_j I(x_i,y_j) \Delta x^2 W(\vert x - x_i \vert,\vert y-y_j \vert, h(x_i,y_j)), \label{eq:Ismooth}
\end{equation}
where $i$ and $j$ sum over the number of pixels in the $x$ and $y$ directions, $\Delta x$ is the pixel spacing, and $W$ is a 2D smoothing kernel. We then implemented an adaptive smoothing by relating $h$ to the local pixel intensity using
\begin{equation}
h(x,y) = h_{\rm min} \sqrt{\frac{I_{\rm max}}{<I(x,y)>}}, \label{eq:hfac}
\end{equation}
where $h_{\rm min}$ is the beam size at $I = I_{\rm max}$. Since both the smoothed intensity and the smoothing length itself are mutually dependent, we solved (\ref{eq:hfac}) and (\ref{eq:Ismooth}) iteratively, according to the usual practice in SPH \citep[e.g.,][]{Price2007b}. We adopted the usual cubic B-spline kernel \citep{Monaghan1985} for the interpolation. We also employed the exact sub-pixel interpolation method from \citet{Petkova2018} to ensure that the total flux is conserved by the interpolation procedure. The new set of pixels $(x,y)$ is arbitrary, but in practice we interpolated to a set of pixels that are the same as those in the original image.

\begin{figure}[htb]
   \includegraphics[width=\columnwidth]{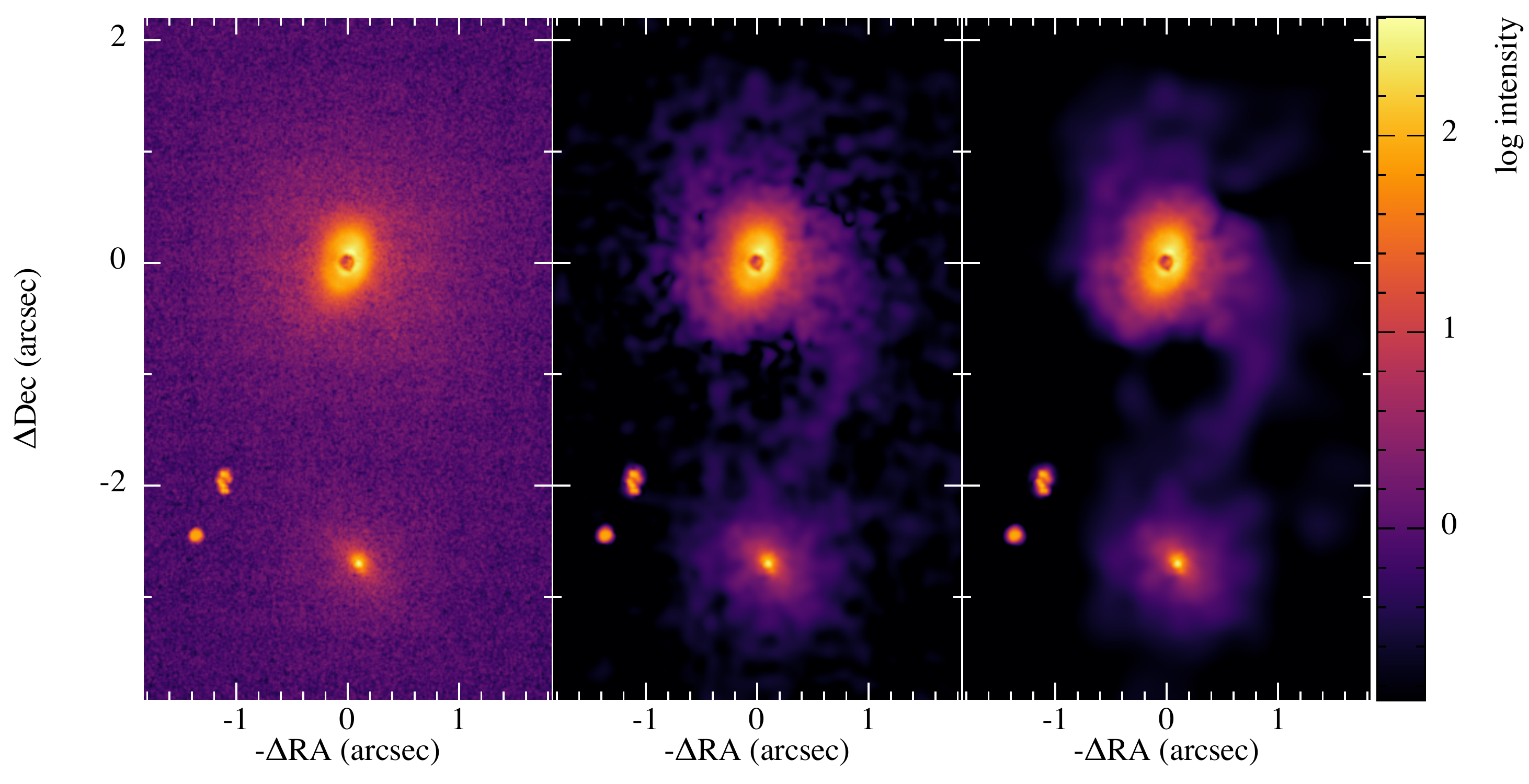}
    \caption{J-band polarized intensity images. {\bf Left panel:} Native J-band image shown in log stretch. {\bf Middle panel:} 
    De-noised J-band polarized intensity image where the Stokes Q and U images were combined before applying the de-noising procedure. {\bf Right panel:} De-noised J-band image where the Q and U images were de-noised separately before combining to produce the polarized intensity de-noised image.}%
    \label{fig:denoise-compare}
\end{figure}

We chose $I_{\rm max}$ to be the maximum intensity in the image and found best results with $h_{\rm min} = 1.25$ times the original pixel spacing for the UX Tau J-band polarized intensity image presented in Fig.~\ref{fig:denoised}, right panel. For polarized intensity images we found it best to de-noise each polarization separately before combining to form a total polarized intensity image. This is shown in fig.~\ref{fig:denoise-compare}. A paper describing our open source de-noising tool in detail is in preparation (Price et al. 2020, in prep).

\end{appendix}
\end{document}